\documentclass[reprint,amssymb,pra]{revtex4-1}
\usepackage[utf8]{inputenc}
\usepackage[T1]{fontenc}
\usepackage{graphicx}
\usepackage{amsmath}
\usepackage{amssymb}
\usepackage{mathrsfs}
\usepackage{braket}
\usepackage{geometry}
\usepackage{here}
\usepackage{lmodern}
\usepackage{array}
\usepackage{natbib}
\usepackage{url}
\usepackage{textcase}
\usepackage{bm}
\usepackage{natbib}
\usepackage{color}
\begin{document}
\title{Coherent control of reactive scattering at low temperatures:  
Signatures of quantum interference in the differential cross sections for F~+~H$_2$ and F~+~HD }

\author{Adrien Devolder$^{1}$, Timur Tscherbul$^{2}$, and Paul Brumer$^{1}$}

\affiliation{$^{1}$Chemical Physics Theory Group, Department of Chemistry, and Center for Quantum Information and Quantum Control, University of Toronto, Toronto, Ontario, M5S 3H6, Canada\\
$^{2}$Department of Physics, University of Nevada, Reno, NV, 89557, USA}

\date{\today}

\begin{abstract}
Fundamental entanglement related challenges have prevented
quantum interference-based control (i.e. coherent control)
of collisional cross sections from being implemented in the laboratory.
Here, differential cross sections for reactive scattering at low temperatures 
 are shown to provide a unique opportunity to display such
interference-based control by forming coherent superpositions of degenerate rotational states of  reactant molecules $|jm\rangle$ with different $m$. In particular, we identify and quantify a unique
signature of coherent control in reactive scattering
with applications to F~+~H$_2 \rightarrow$ H + HF and HF + D $\leftarrow$ 
F + HD $\rightarrow$ HD + F at $11$~K. Control is shown to be extensive.
\end{abstract}

\maketitle

Coherent control of atomic and molecular processes (for a review until
2012, see \cite{Brumer_book}), i.e., the
use of quantum interference to effect molecular outcomes, has
proven enormously successful for certain classes of processes. 
These include assorted processes, such as light-induced control of unimolecular
processes such as photodissociation \cite{Sheehy1995}, photoionization \cite{Zhu1995},		      control of currents in live brain cells \cite{Lavigne:20}, control of population
transfer between system eigenstates \cite{Kleiman1995a}, control of internal conversion\cite{Grinev2015},  etc. By preparing
multiple interfering pathways as initial states, primarily by laser
excitation, quantum-interference-based control over various processes,
has been demonstrated both computationally (e.g., \cite{Brumer_book,Brumer1986a,Brumer1986b,Tannor1985}) and experimentally (e.g., \cite{Sheehy:95,Zhu:95,Shnitman:96}).
However, control over the wide class
of collisional processes such as chemical reactions requires, in general, entanglement
between the translation motion of the colliding partners and their
internal degrees of freedom \cite{Shapiro1996,Brumer_book, Omiste2018}, a non-trivial experimental challenge
for molecular collisional systems of interest.
While this requirement can be relaxed for coherent 
control over the	 differential cross section (DCS) for A~+~BC collisions by forming coherent superpositions of  energetically degenerate states of BC,  coherent control of reactive scattering is yet to be demonstrated experimentally.

Such a demonstration would be particularly valuable for cold and ultracold chemical reactions in the quantum regime \cite{Krems:08}, which have become amenable to experimental studies owing to recent  advances in cooling,  trapping, and manipulating molecular gases \cite{Bohn:17}. These studies have  revealed a number of fascinating phenomena, such as resonant scattering in cold He$^*$~+~H$_2$ \cite{Henson:12,Klein:16} and He~+~NO \cite{Vogels:15} collisions, stereodynamics of H$_2$~+~HD collisions at 1 K \cite{Perreault2017,Perreault:18}, quantum tunneling in the chemical reaction F~+~H$_2$ $\to$ HF~+~H at cold temperatures\cite{Tizniti2014},  and electric field control of the chemical reaction 2KRb $\to$ K$_2$ + Rb$_2$ at 50~nK \cite{Ni:10,Miranda:11}. Several theoretical studies explored the  effects of molecular polarization \cite{Aldegunde:06} and alignment  \cite{Croft:18,Jambrina:19} on ultracold collision dynamics.
 
 There are three primary motivations for using coherent control to manipulate cold molecular collisions. First, in the low-temperature regime, the number of quantum states of the reactants (including partial waves for the relative motion) is dramatically reduced \cite{Krems:08,Bohn:17}, minimizing thermal fluctuations and decoherence, and thereby enhancing quantum controllability of molecular processes. Second, because coherent control relies  on the very general phenomenon of quantum interference, it could potentially be applied to a much wider range of molecular species than dc field control, which typically employs, e.g., Feshbach resonances to tune the scattering properties of ultracold atoms and molecules \cite{Chin:10}. Thus, coherent control may prove advantageous in experimental settings, where the presence of dc fields can cause  undesirable perturbations, such as in precision measurements using atomic and molecular clocks \cite{Derevianko:11,Ludlow:15}.
 Third, recent experimental studies of low-temperature collisions of molecules in single rotational states \cite{Ni:10,Vogels:15,Tizniti2014,Miranda:11} and in superpositions thereof  \cite{Perreault2017,Perreault:18} now provide
an experimental platform where such quantum interference-based  control  of reactive scattering can be carried out.

Here we computationally demonstrate that (a) extensive control
can be achieved over the DCS for reactive F~+~H$_2 \rightarrow$
HF~+~H, and F~+~HD $\rightarrow$ HF~+~D or DF~+~H in cold (11~K) scattering,
by preparing a superposition of magnetic sublevels of {\it ortho}-H$_2$ and HD, and
(b) that such control displays a unique measurable signature, readily 
identifying quantum interference as the basis for control. Computational results are
obtained within a scenario that can be realized in modern experiments using merged beams of H$_2$ molecules created in superpositions of rotational states using, e.g., Stark-induced adiabatic Raman passage (SARP)  \cite{Perreault2017,Perreault:18}. The results also provide motivation for the extension to systems of particular interest in
ultracold chemistry, such as those involving alkali-metal dimers KRb, NaK, and NaLi \cite{Ospelkaus:10,Ni:10,Park:15,Rvachov:17}, and $^2\Sigma$ molecular radicals CaF, SrF, and SrOH \cite{Barry:14,McCarron:18,Cheuk:18,Anderegg:18,Kozyryev:17}.

We propose to control a reactive atom-molecule collision A~+~BC~$\to$~AB~+~C by preparing an initial state $\ket{\psi_{s}}$ as a superposition of two $m$-states of a diatomic molecule,
a procedure that has been experimentally demonstrated using 
SARP \cite{Mukherjee2011,Mukherjee2014,Perreault2017}
\begin{equation}
\ket{\psi_{s}}=\cos(\eta)\ket{1,v,j,m_1}+\sin(\eta)\exp(i\beta)\ket{1,v,j,m_2}
\label{eq.sup}
\end{equation}
Here $v$ and $j$ are the vibrational and rotational quantum numbers, respectively, and the first quantum number  $\alpha$ corresponds to the chemical arrangement (i.e. A~+~BC, B~+A~C or 
C~+~AB) with $\alpha=1$ denoting the initial A~+~BC arrangement. The parameter $\eta$ sets
the relative population of the two states in the superposition with relative phase $\beta$.

Given this initial superposition state (\ref{eq.sup}), the
DCS to a final state
($\alpha',v',j',m'_j$) can be written as the sum of two terms:
\begin{equation}
%\frac{d}{d\Omega}(\theta,\phi)=
\sigma_{s \rightarrow \alpha' v' j' m'_j}(\theta,\phi)=
{\sigma^{incoh}_{s \rightarrow \alpha' v' j' m'_j}} (\theta)+{\sigma^{int}_{s \rightarrow \alpha' v' j' m'_j}}(\theta,\phi),
\label{eq:diff_sup}
\end{equation}
where the subscript $s$ denotes the initial superposition state [Eq.(\ref{eq.sup})].
The first of these terms, ${\sigma^{incoh}_{s \rightarrow \alpha' v' j' m'_j}}(\theta)$ 
is an incoherent contribution equivalent to the DCS from a mixture
of $m$ states with probabilities $\cos^2(\eta)$ and $\sin^2(\eta)$:
\begin{align}\notag
{\sigma^{incoh}_{s \rightarrow \alpha' v' j' m'_j}}(\theta) &= \cos^2(\eta) |f_{1 v j m_1 \rightarrow \alpha' v' j' m'_j}(\theta,\phi)|^2 \\ &+
\sin^2(\eta)|f_{1 v j m_2 \rightarrow \alpha' v' j' m'_j}(\theta,\phi)|^2
\label{eq:diff_incoh}
\end{align}
where $f_{1 v j m_i \rightarrow \alpha' v' j' m'_j}(\theta,\phi,E)$ ($i=1,2$)  are the scattering amplitudes into
product states $|\alpha'v'j',m_j'\rangle$ that are the key quantities to compute:
% The atom-molecule scattering amplitudes are given by:
	\begin{multline}
	f_{\alpha v j m_j \rightarrow \alpha' v' j' m'_j}(\theta,\phi) = \frac{i\pi^{1/2}}{(k_{\alpha v j}k_{\alpha' v' j'})^{1/2}} \\ \times \sum_{J,M}\sum_{\ell,\ell'}\sum_{m'_\ell}i^{\ell-\ell'}  (2\ell+1)^{1/2} \begin{bmatrix}
	j & \ell & J \\ m_j & 0 & M
	\end{bmatrix}\begin{bmatrix}
	j' & \ell' & J \\ m'_j & m_\ell'& M
	\end{bmatrix}\\  \times\left[\delta_{\alpha \alpha'}\delta_{vv'}\delta_{jj'}\delta_{\ell\ell'}- S_{\alpha v j m_j \rightarrow \alpha' v' j' m'_j}^J(E)\right]Y_{\ell'm'_\ell}(\theta,\phi).
	\label{eq:scatt_amp}
	\end{multline}
Here, $\ell$($\ell'$) is the initial (final) partial wave and
$m_\ell$($m'_\ell$) is the initial (final) projection of $\vec{\ell}$ on
the space-fixed quantization axis $Z$, $E$ is the collision energy, $k_{\alpha v
j}$($k_{\alpha' v' j'}$) are the initial (final)
relative momenta.
%, related to the relative kinetic energy before (after) the collision. 
The symbols in brackets are the
Clebsh-Gordan (CG) coefficients, $Y_{\ell'm'_\ell}(\theta,\phi)$
are the spherical harmonics and $S_{\alpha v j m_j
\rightarrow \alpha' v' j' m'_j}^J(E)$ are the S-matrix elements.

The second term in Eq.~{(\ref{eq:diff_sup})}, ${\sigma^{int}_{s \rightarrow \alpha'
v' j' m'_j}}$, is the interference
contribution, in which coherent control is manifest. 
Specifically, coherent control occurs via the quantum interference
between the two scattering pathways arising from the initial
$m$-superposition (\ref{eq.sup}). The interference term is then
given by:
%The
%principle of coherent control is tuning the value of this term
%via a variation of the parameters ($\eta$ and $\beta$) of the
%initial preparation. 
\begin{multline}
%\begin{split}
{\sigma^{int}_{s \rightarrow \alpha' v' j' m'_j}} (\theta,\phi)=\cos(\eta) \sin(\eta) \\ \times [e^{-i\beta} f_{1 v j m_1 \rightarrow \alpha' v' j' m'_j}(\theta,\phi) f_{1 v j m_2 \rightarrow \alpha' v' j' m'_j}^*(\theta,\phi) \\ + e^{i\beta} f_{1 v j m_1 \rightarrow \alpha' v' j' m'_j}^*(\theta,\phi) f_{1 v j m_2 \rightarrow \alpha' v' j' m'_j}(\theta,\phi) ].
%\end{split}
\label{eq:diff_int}
\end{multline}
Note the characteristic difference between the direct terms [Eq. (\ref{eq:scatt_amp})]
and the interference term [Eq. (\ref{eq:diff_int})]. Specifically,  the latter shows a dependence
of scattering on the azimuthal angle $\phi$ whereas the direct terms are $\phi$-independent.
Hence, \textit{$\phi$-dependent scattering is a unique signature of interfering
quantum pathways in the differential scattering cross section}. Previous experimental studies \cite{Perreault2017,Perreault:18}
on scattering with this type of superposition state did not measure this dependence insofar as detection included averaging over $\phi$.

The $\phi$-dependence in Eq. (\ref{eq:diff_int}) originates from the $e^{im'_\ell\phi}$ term in the 
spherical harmonics contribution to the
scattering amplitudes [see Eq. (\ref{eq:scatt_amp})]. The
two CG coefficients in Eq. (\ref{eq:scatt_amp}) ensure that
$m'_\ell = m_j-m'_j$. Consequently, 
 the scattering amplitudes can be written as 
$e^{i(m_j-m'_j)\phi}$ times a $\theta$-dependent part
\begin{multline}
f_{\alpha v j m_j \rightarrow \alpha' v' j' m'_j}(\theta,\phi)=e^{i(m_j-m'_j)\phi} |f_{\alpha v j m_j \rightarrow \alpha' v' j' m'_j} (\theta)| \\ \times e^{i\xi_{\alpha v j m_j \rightarrow \alpha' v' j' m'_j}(\theta)}
\label{eq:scatt_amp_polar}
\end{multline}
The products of the two scattering amplitudes in Eq. (\ref{eq:diff_int}) are then
proportional to $e^{i(m_1-m_2)\phi}$. Inserting Eq. (\ref{eq:scatt_amp_polar})
into Eq. (\ref{eq:diff_int}) gives the DCS for a final
state as:
\begin{multline}
%\begin{split}
\sigma_{s\rightarrow \alpha' v' j' m'_j}(\theta,\phi)= {\sigma^{incoh}_{s \rightarrow \alpha' v' j' m'_j}}(\theta)+2 \cos(\eta) \sin(\eta) \\ \times |f_{\alpha v j m_1 \rightarrow \alpha' v' j' m'_j}(\theta)| \times |f_{\alpha v j m_2 \rightarrow \alpha' v' j' m'_j}(\theta)| \\  \times \cos \bigl{(}(m_1-m_2)\phi+ \Delta\xi(\theta) + \beta \bigr{)}
%\end{split}
\label{eq:diff_phi_state}
\end{multline}
where $\Delta \xi(\theta) =\xi_{\alpha v j m_1 \rightarrow \alpha' v' j' m'_j}(\theta)  -\xi_{\alpha v j m_2 \rightarrow \alpha' v' j' m'_j}(\theta) $ is the  difference in the phase of the scattering amplitudes. 
%and the sum on all internal states of the final arrangement, 
The DCS for a specific final arrangement $\alpha'$, discussed below, is obtained by summing over all rovibrational states in this arrangement:
\begin{multline}
\sigma_{s \rightarrow \alpha'}(\theta,\phi)= \sigma^{incoh}_{s \rightarrow \alpha'}(\theta)
+2 \cos(\eta) \sin(\eta)  F_{s \rightarrow \alpha'}(\theta) \\  \times \cos \bigl{(}(m_1-m_2)\phi+\xi_{s \rightarrow \alpha'}(\theta)+\beta \bigr{)}.
\label{eq:diff_phi}
\end{multline}
where $F_{s \rightarrow \alpha'}(\theta)$ and $\xi_{s \rightarrow \alpha'}(\theta)$ are, respectively, the magnitude and the phase of $\sum_{v',j',m'_j}f_{\alpha v j m_1 \rightarrow \alpha' v' j' m'_j}(\theta,\phi)f^*_{\alpha v j m_2 \rightarrow \alpha' v' j' m'_j}(\theta,\phi)$.

Equation (\ref{eq:diff_phi}) provides the central result of this work, which gives the DCS in terms of the scattering amplitude and the control parameters $\eta$ and $\beta$ of the initial coherent superposition of the reactant molecule's rotational states (\ref{eq.sup}).
Below, we expose the $\theta$ dependence of the cross section via the amplitude $A$, defined as 
\begin{equation}
A(\eta,\theta)=2 \cos(\eta) \sin(\eta) F_{s  \rightarrow \alpha'}(\theta)
\label{eq:Adef}
\end{equation}

Given an initial superposition state [Eq. (\ref{eq.sup})], control can be 
affected by varying the  $\eta$ and $\beta$ parameters.
Note that several structural features of the interference term are universal, i.e. 
independent of the system
under consideration. These include the $e^{i(m_j-m'_j)\phi}$ dependence on 
$\phi$, the role of $\beta$ as a phase in the DCS,
and the dependence on $\eta$ through $\cos(\eta)\sin(\eta)$. Below,
we examine the case of a symmetric superposition ($\eta= \pi/4$), which
gives the maximum value of $\cos(\eta)\sin(\eta)$, and set $\beta=0$ since
variations in $\beta$ just correspond to a shift in phase of the interference
term. We emphasize  below the $\phi$ dependence of the cross sections because it is a unique signature
of quantum interference contribution to the scattering. The extent of this
contribution can be quantified via the visibility $\mathcal{V}(\theta)$ manifest here as 
 the difference between the maximum and minimum DCSs
as a function of $\phi$, scaled by the sum of both
%${\sigma^{max(\phi)}_{s \rightarrow \alpha' v' j' m'_j}}(\theta,\phi),
%{d\sigma^{min(\phi)}_{s \rightarrow \alpha' v' j' m'_j}}(\theta,\phi)$
%of the DCS as a function of $\phi$. Specifically, 
\begin{equation}
\mathcal{V}(\theta) =\frac{ {\sigma^{max(\phi)}_{s \rightarrow \alpha' v' j' m'_j}}(\theta,\phi)-
{\sigma^{min(\phi)}_{s \rightarrow \alpha' v' j' m'_j}}(\theta,\phi)}
% \\ \times
{{\sigma^{max(\phi)}_{s \rightarrow \alpha' v' j' m'_j}}(\theta,\phi)+
{\sigma^{min(\phi)}_{s \rightarrow \alpha' v' j' m'_j}}(\theta,\phi)}.
 \label{eq:visibility}
 \end{equation}
 From Eqs.~(\ref{eq:diff_phi}) and (\ref{eq:Adef}), we obtain the visibility as the ratio $A(\theta)/\sigma^{incoh}_{s \rightarrow \alpha'}(\theta)$. Indeed, the extent to which coherent control is significant is dictated by the relative
magnitude of the amplitude $A$ and the incoherent contributions, i.e. by the
visibility $\mathcal{V}$ of the interference fringes. 
For the case of two pathways, $\mathcal{V}$ satisfies \cite{Scholak:17,Qureshi:19}.
\begin{equation}
\mathcal{P}(\theta)^2 + \mathcal{V}(\theta)^2 \le 1,
\label{eq:V}
\end{equation}
an expression of wave-particle duality, with $\mathcal{V}$ measuring the wave-like behavior and $\mathcal{P}(\theta)$ is the path distinguishability\cite{Qureshi:19}.  The highest value for the visibility $\mathcal{V}$ is then 1, corresponding to a situation where the two pathways are completely indistinguishable. The closer it is to unity, the greater the contribution of the
interfering indistinguishable pathways to the scattering. Note that the positivity of the DCS implies  that  $A\le \sigma^{incoh}_{s \rightarrow \alpha'}(\theta)$, which is consistent with the unitary limit for the visibility.

We now apply the methodology developed above to 
explore the possibility of controlling cold chemical reactions F~+~H$_2 \rightarrow$ HF~+~H and F~+~HD~$\rightarrow$ H + DF and D + HF. Recent experimental advances in preparing coherent superpositions of rotational states of H$_2$  \cite{Perreault2017,Perreault:18} and in reactive
scattering of F + H$_2$ at 11~K \cite{Tizniti2014} indicate the feasibility of an
experiment (the first coherent control experiment of reactive scattering) of the
kind motivated by the results below. The reaction DCS are computed using Eqs.  	(\ref{eq:diff_sup}-\ref{eq:diff_int}) parametrized by the $S$-matrix elements obtained via a numerically exact time-independent quantum reactive scattering approach \cite{Skouteris2000} (see the Supplemental Material \cite{SM} for further details). After, for a specific arrangement, we summed over all rovibrational states in this arrangement.  
Similar control results were obtained at 1~K. We note that while the Stark-Werner PES used here is known to overestimate the true F~+~H$_2$ reaction rate by a factor of $\simeq$3 at 11~K  \cite{Tizniti2014},  it does provide a qualitatively correct picture of this tunneling-dominated reaction \cite{Balakrishnan2001,Zhu:02}, and can therefore be used to establish qualitative trends in controlling low-temperature reaction rates, the main goal of this work.

%[ADRIEN: ARE THESE ULTRA-COLD RESULTS RELIABLE, OR WOULD WE NEED BETTER POTENTIAL SURFACES?]
%for the chemical reactions  F~+~H$_2$ and F~+~HD 

%As an initial superposition state of the reactant molecule, we choose 
We consider two different $m$-superpositions of the magnetic sublevels of the first excited rotational  state
($j=1$) of the vibrational ground state ($v=0$) of either
{\it ortho}-H$_2$ or HD. The first superposition, between $m_1=-1$ and
$m_2=0$ , is denoted -1/0 and the second, between $m_1=-1$ and $m_2=+1$, is
denoted -1/+1.
% ( \textcolor{blue}{see the Supplemental Material for more results}.
%Of the numerous supportive results obtained, several are reported below, with some additional computations in the

\begin{figure}%{ht}
% \begin{figure}[H]
	\centering
	\includegraphics[width=\columnwidth]{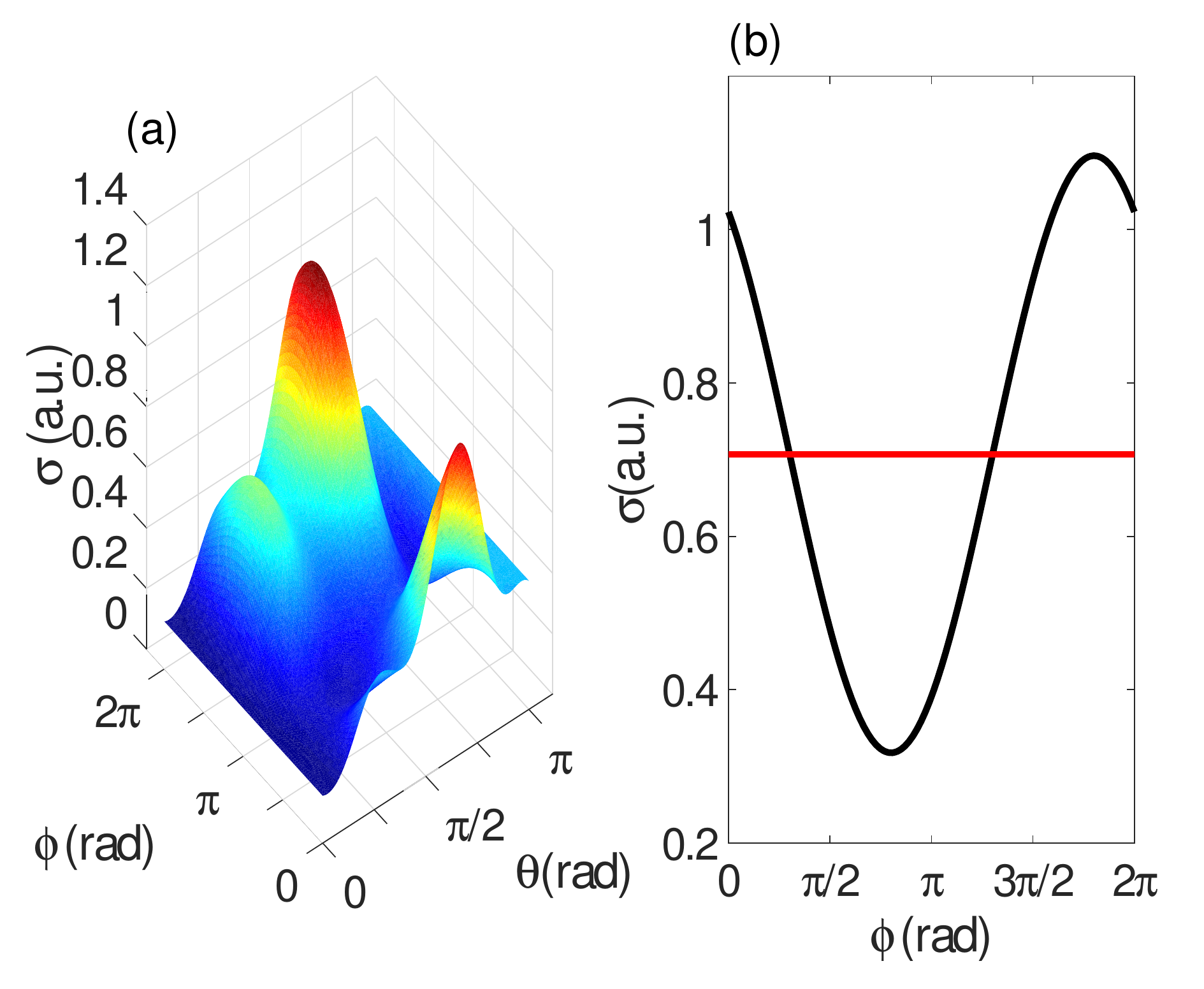}
	\caption{Reactive DCS for F~+~H$_2$ with the initial superposition -1/0 ($\eta=\pi/4$ and $\beta=0$). (a) 3D plot representing the $\theta$- and $\phi$-dependences of the DCS. (b) $\phi$-dependence at fixed $\theta$ ($\theta=0.65$ $\pi$ rad). The DCS is plotted in black while the incoherent contribution is plotted in red.}
	\label{fig1}
\end{figure}

The calculated DCS for the F~+~H$_2(-1/0) \rightarrow$ HF + H reaction is shown in Fig.~\ref{fig1}(a) as a function of $\theta$ and $\phi$. Although the interference
contribution affects both the dependence on $\theta$ and $\phi$,
the explicit $\phi$ dependent signature of interference at the $\theta$ value, for which the DCS is maximal, 
(here, $\theta = 0.65 \pi$), is shown in Fig.~\ref{fig1}(b). The  $\phi$ dependence of the DCS is seen to oscillate  about the direct term, with a frequency equal to the difference of the projections of magnetic sublevels in the superposition (\ref{eq.sup}), as anticipated from Eq. (\ref{eq:diff_phi}). A simple computation gives $\mathcal{V} = 0.54$, whose square, $\approx$0.33, constitutes a 
considerable interference contribution.

\begin{figure}
	\centering
	\includegraphics[width=\columnwidth]{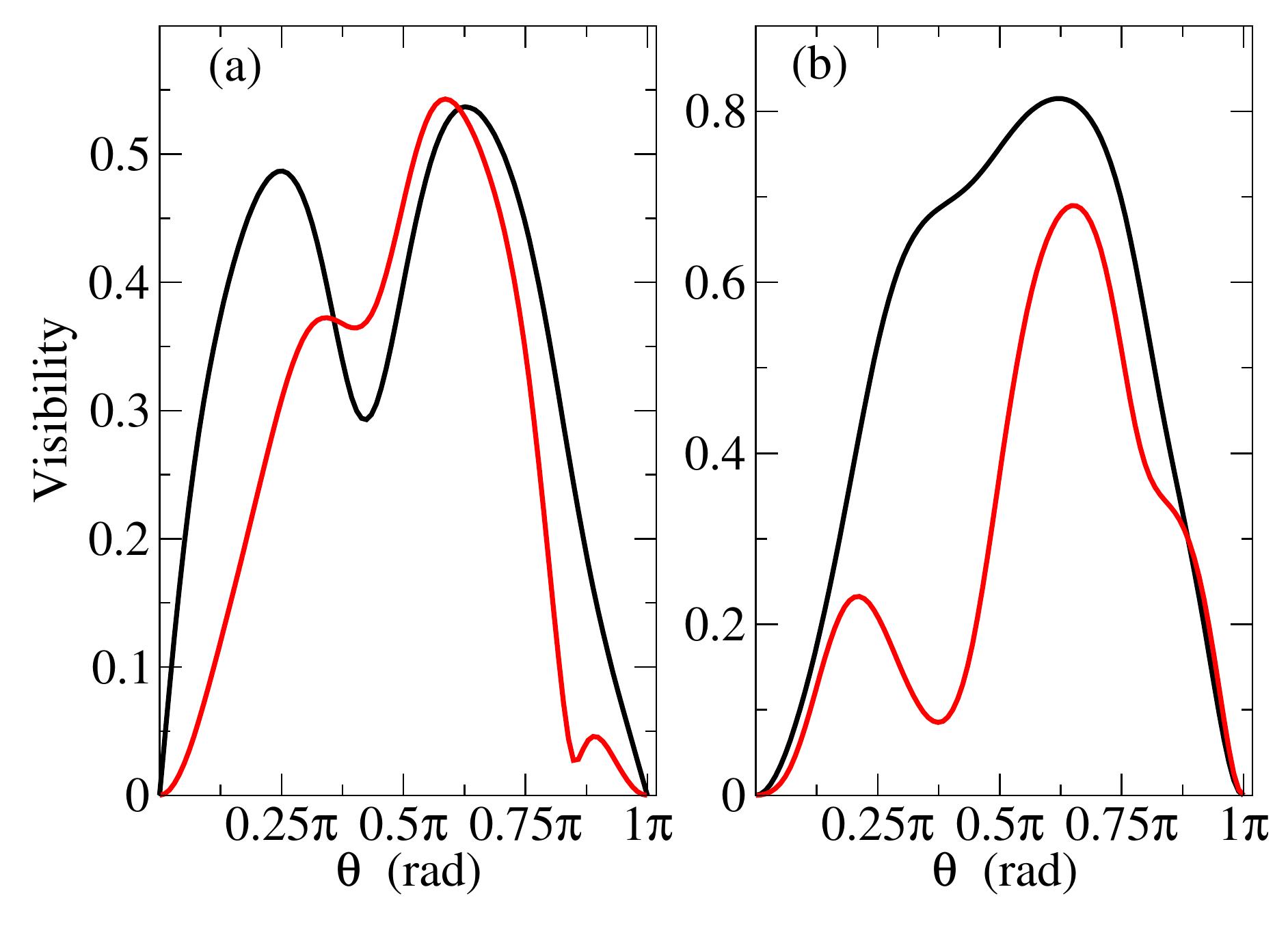}
	\caption{$\theta$-dependence of the visibility for (a) F~+~H$_2 \rightarrow$ HF~+~H with the initial superpositions -1/0 (black) and -1/+1 (red), and for (b)  F~+~HD $\rightarrow$ DF~+~H (black) and HF~+~D (red) with the initial superpositions -1/+1.}
	\label{Vis}
\end{figure}

Whereas the functional form (albeit not the magnitude) 
of the $\phi$ dependence of the DCS given by $\exp{[i(m-m')\phi]}$
is universal, the $\theta$ dependence of the amplitude $A$ (\ref{eq:Adef}), of the incoherent contribution (\ref{eq:diff_incoh}) and therefore of the visibility (\ref{eq:visibility}) are system dependent. 
Figure~\ref{Vis} (a) shows the $\theta$ dependence of the visibility for the 
F + H$_2$ superpositions -1/0 and -1/+1. No clear advantage of one superposition over the other appears, with the maximal visibility $\mathcal{V}_{max}=0.54$. The interference clearly vanishes in the forward and backward directions,  where the scattering amplitudes are zero.

Similarly to the amplitude $A(\theta)$, the phase $\xi_{s\to \alpha'}(\theta)$ of DCS oscillations (\ref{eq:diff_phi}), given by the difference of phases of scattering amplitudes, is system-dependent. The phase is  particularly sensitive to the presence of a symmetry between the  different components of the superposition (\ref{eq.sup}). For example, the two states of the superposition -1/+1 are related by the time-reversal symmetry and the phase $\xi$ of the DCS oscillations is zero for all $\theta$. Inversely, for the superposition -1/0, we observe a $\theta$-dependent phase $\xi(\theta)$ of the DCS oscillations [see, e.g., Fig.~\ref{fig1}(b)]. Thus, by measuring the angular dependence of the DCS for molecules reacting in quantum superposition states (\ref{eq.sup}), it is possible to infer information not only about the magnitude, but also about the phase of the scattering amplitudes.

\begin{figure}
	\centering
	\includegraphics[width=\columnwidth]{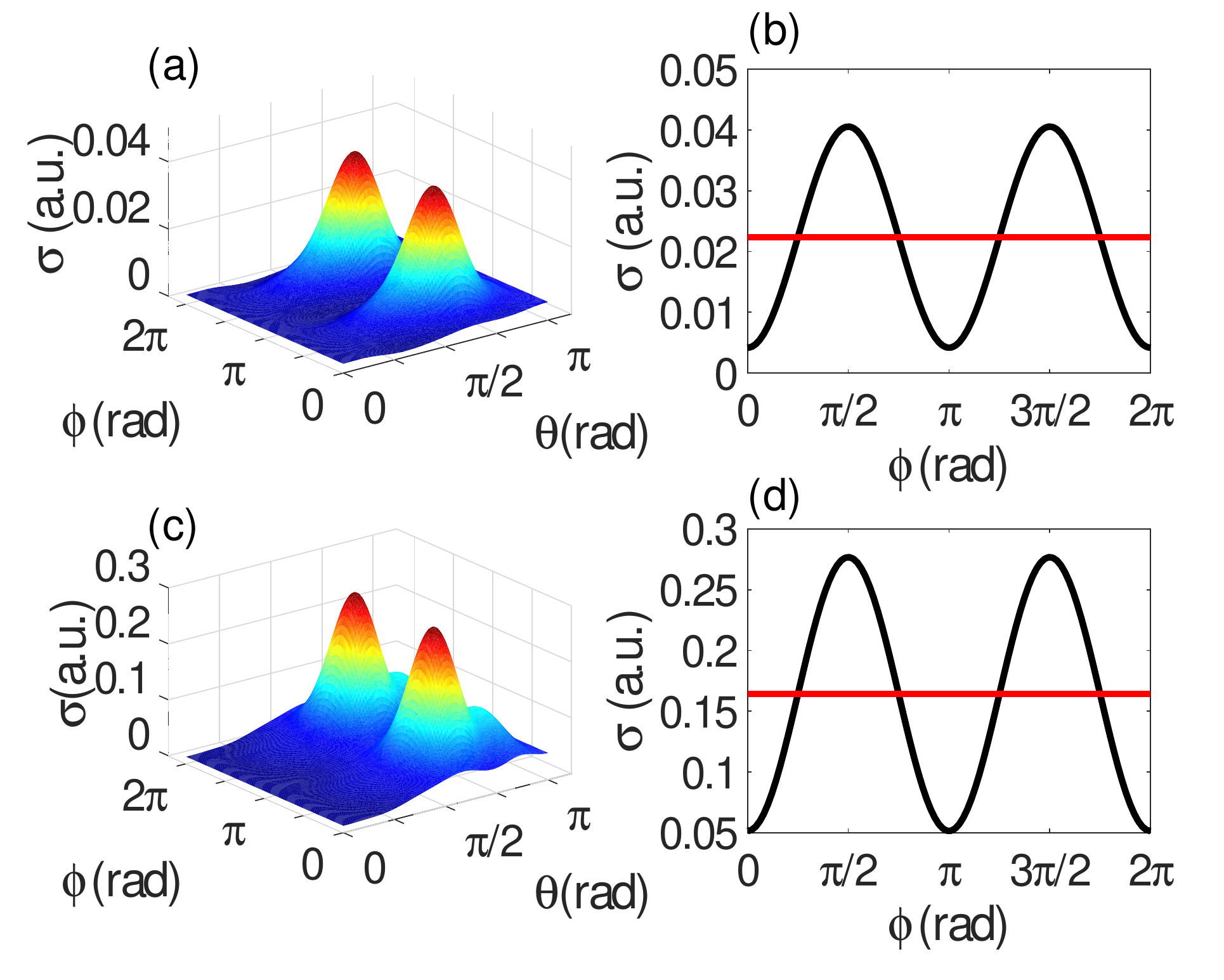}
	\caption{Reactive DCS for F~+~HD with the initial superposition 
	 -1/+1 ($\eta=\pi/4$ and $\beta=0$). (a) and (c): 3D plot of the $\theta$- 
	 and $\phi$-dependent DCS. (b) and (d): $\phi$-dependence of the DCS 
	  at  $\theta=0.64$ $\pi$. Panels (a) and (b) correspond to the reactive channel
	   DF~+~H while panels (c) and (d) correspond to the reactive channel HF~+~D.  The DCS are plotted in blue while the incoherent contributions are plotted in red.
	   Note the difference in ordinate scale for the two different products.}
	\label{fig4}
\end{figure}

%========================figure 4 indidual F + HD ==========
Control over reactive F~+~HD scattering presents additional challenges and opportunities,
insofar as successful control should be able to 
\textit{selectively distinguish} between the H~+~FD and D~+~HF product channels. Figure \ref{fig4} shows the results for 
controlling the individual DCSs for each of the product channels with the initial superposition -1/+1. Extensive $\phi$-dependences for the both reactive channels are evident with large visibility, $\mathcal{V} = 0.81$ for
the H + DF channel, and 0.69 for the H + FD channel. Note that the DCS is
larger for D~+~HF than for the DF~+~H channel, a consequence of larger tunneling for the hydrogen
atom. 

\begin{figure}
	\centering
	\includegraphics[width=\columnwidth]{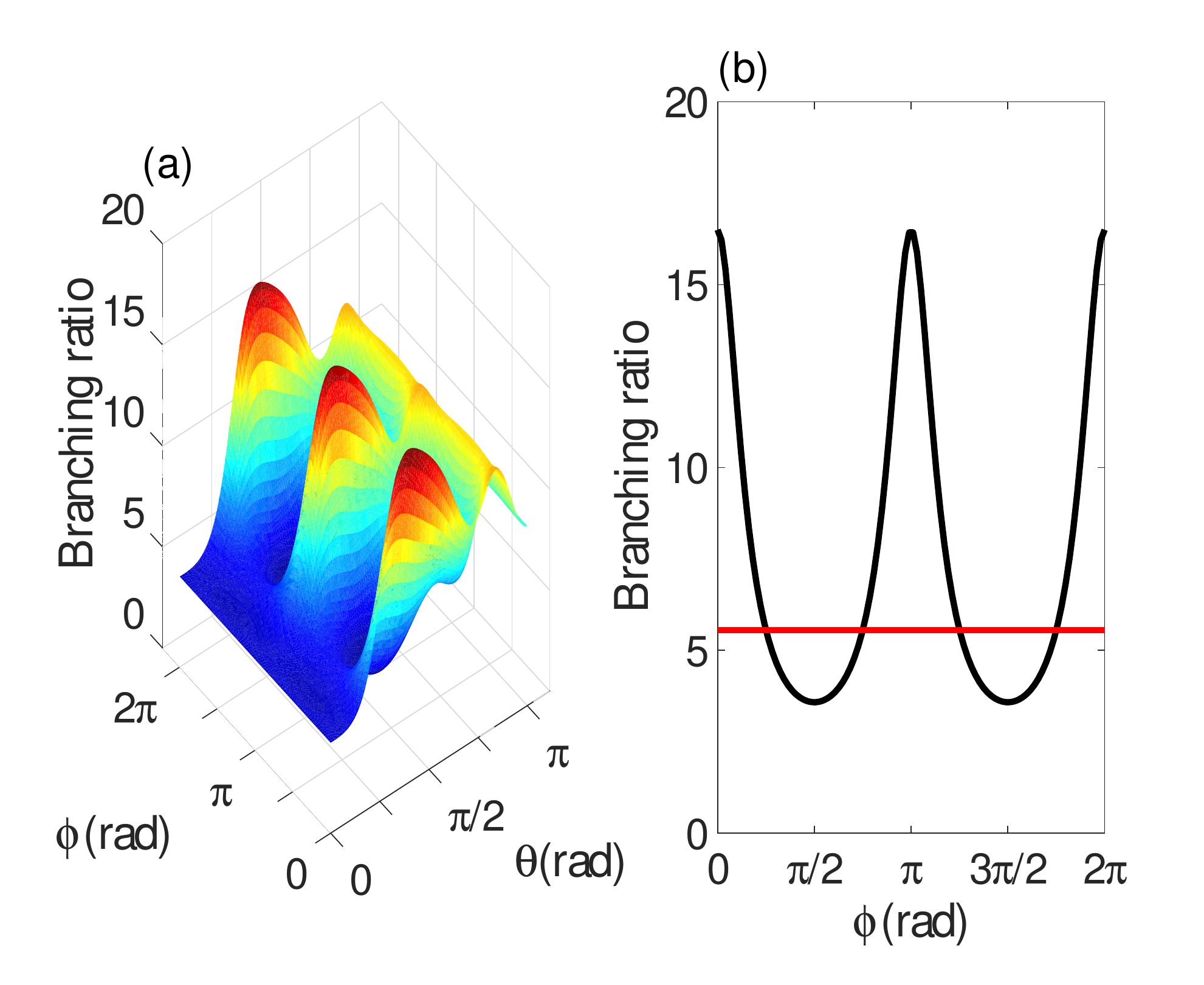}
	\caption{Differential branching ratio (HF~+~D)/DF~+~H) of the DCS 	
	between the reactive channels
	 for F+HD with the initial superposition -1/1. (a) 3D surface plot of the $\theta$- and 
	 $\phi$-dependences of the DCS. (b) $\phi$-dependence  of the branching ratio at a fixed
	  $\theta=0.47\pi$. The differential branching ratio is plotted in blue while the incoherent contribution is plotted in red.}
	\label{fig14}
\end{figure}

The vanishing of the phases $\xi(\theta)$ for both of the product channels [see Figs.~\ref{fig4}(b) and (d)] due to the time-reversal symmetry implies that the selectivity in the control over these channels can only arise from differences in the $\theta$-dependence of the visibilities $\mathcal{V}$ shown in Fig.~\ref{Vis}(b). We observe that the best selectivity is expected for $\theta=\pi/4{-}\pi/2$ rather than at the maximum of the visibilities, where the $\theta$-dependence is similar for both of the product channels. To further reveal this selectivity, we plot in Fig.~\ref{fig14} the differential branching ratio, the ratio of the DCS $\sigma(\theta,\phi)$ for the two product channels HF~+~D  and DF~+~H.
%The left hand panel shows the ratio of the DCS for the two products.  
The $\phi$-dependence of the ratio shown in Fig.~\ref{fig14}(b) 
 is seen to vary  over a wide range, from 3.6 to 16.5, clearly demonstrating interference-based
\textit{product channel selectivity} with interference quantified at $\mathcal{V} =0.64$. We note that since the $\phi$-dependence of the HF+D/DF+H branching ratio is given by the ratio of two in-phase oscillations, it is not symmetric about the ratio of incoherent contributions shown by the horizontal line in Fig.~\ref{fig14}(b).

In summary, we have computationally demonstrated extensive
quantum-interference based coherent control over low-temperature differential scattering in the F~+~H$_2$ and F~+~HD chemical reactions using an approach that is experimentally feasible.
Quantum interference control is manifest explicitly in a signature experimental
observable, the non-zero $\phi$ dependence of the scattering.   
The successful experimental demonstration of this control would
open the entire class of scattering processes to coherent control.

The proposed control scenario is completely general and can be extended to all reactive scattering processes of interest to ultracold chemistry. Significantly, this scheme allows for controlling chemical reactions that are not readily susceptible to traditional electric and magnetic field control, such as those involving homonuclear alkali-metal dimers \cite{Staanum:06,Zahzam:06,Viteau:08,Quemener:12} and H$_2$, the most abundant molecule in the Universe and one of the very few molecules, whose reaction dynamics can be studied theoretically with spectroscopic accuracy \cite{Althorpe:03,Ren:08,Zhang:16,Klein:16}.
Furthermore, because $\phi$-dependent interference 
is observed only if the atom-molecule PES is anisotropic, %\cite{Footnote}  
this dependence provides direct insight into the angular dependence of the PES. Measuring the $\phi$-dependent DCS in a coherently controlled cold  scattering experiment could, therefore, serve a useful probe of the interaction anisotropy in atom-molecule reactive scattering.

This work was supported by the U.S. Air Force Office for Scientific Research (AFOSR)
under Contract No. FA9550-19-1-0312. Discussions with
Prof. Millard Alexander, University of Maryland, are gratefully 
acknowledged. We thank Dario DeFazio for providing his test code for the calculation of DCS.


\begin{thebibliography}{53}%
\makeatletter
\providecommand \@ifxundefined [1]{%
 \@ifx{#1\undefined}
}%
\providecommand \@ifnum [1]{%
 \ifnum #1\expandafter \@firstoftwo
 \else \expandafter \@secondoftwo
 \fi
}%
\providecommand \@ifx [1]{%
 \ifx #1\expandafter \@firstoftwo
 \else \expandafter \@secondoftwo
 \fi
}%
\providecommand \natexlab [1]{#1}%
\providecommand \enquote  [1]{``#1''}%
\providecommand \bibnamefont  [1]{#1}%
\providecommand \bibfnamefont [1]{#1}%
\providecommand \citenamefont [1]{#1}%
\providecommand \href@noop [0]{\@secondoftwo}%
\providecommand \href [0]{\begingroup \@sanitize@url \@href}%
\providecommand \@href[1]{\@@startlink{#1}\@@href}%
\providecommand \@@href[1]{\endgroup#1\@@endlink}%
\providecommand \@sanitize@url [0]{\catcode `\\12\catcode `\$12\catcode
  `\&12\catcode `\#12\catcode `\^12\catcode `\_12\catcode `\%12\relax}%
\providecommand \@@startlink[1]{}%
\providecommand \@@endlink[0]{}%
\providecommand \url  [0]{\begingroup\@sanitize@url \@url }%
\providecommand \@url [1]{\endgroup\@href {#1}{\urlprefix }}%
\providecommand \urlprefix  [0]{URL }%
\providecommand \Eprint [0]{\href }%
\providecommand \doibase [0]{http://dx.doi.org/}%
\providecommand \selectlanguage [0]{\@gobble}%
\providecommand \bibinfo  [0]{\@secondoftwo}%
\providecommand \bibfield  [0]{\@secondoftwo}%
\providecommand \translation [1]{[#1]}%
\providecommand \BibitemOpen [0]{}%
\providecommand \bibitemStop [0]{}%
\providecommand \bibitemNoStop [0]{.\EOS\space}%
\providecommand \EOS [0]{\spacefactor3000\relax}%
\providecommand \BibitemShut  [1]{\csname bibitem#1\endcsname}%
\let\auto@bib@innerbib\@empty
%</preamble>
\bibitem [{\citenamefont {Shapiro}\ and\ \citenamefont
  {Brumer}(2012)}]{Brumer_book}%
  \BibitemOpen
  \bibfield  {author} {\bibinfo {author} {\bibfnamefont {M.}~\bibnamefont
  {Shapiro}}\ and\ \bibinfo {author} {\bibfnamefont {P.}~\bibnamefont
  {Brumer}},\ }\href@noop {} {\emph {\bibinfo {title} {Quantum Control of
  Molecular Processes}}}\ (\bibinfo  {publisher} {Wiley-VCH},\ \bibinfo {year}
  {2012})\BibitemShut {NoStop}%
\bibitem [{\citenamefont {Sheehy}\ \emph
  {et~al.}(1995{\natexlab{a}})\citenamefont {Sheehy}, \citenamefont {Walker},\
  and\ \citenamefont {DiMauro}}]{Sheehy1995}%
  \BibitemOpen
  \bibfield  {author} {\bibinfo {author} {\bibfnamefont {B.}~\bibnamefont
  {Sheehy}}, \bibinfo {author} {\bibfnamefont {B.}~\bibnamefont {Walker}}, \
  and\ \bibinfo {author} {\bibfnamefont {L.~F.}\ \bibnamefont {DiMauro}},\
  }\href@noop {} {\bibfield  {journal} {\bibinfo  {journal} {Phys. Rev. Lett.}\
  }\textbf {\bibinfo {volume} {74}},\ \bibinfo {pages} {4799} (\bibinfo {year}
  {1995}{\natexlab{a}})}\BibitemShut {NoStop}%
\bibitem [{\citenamefont {Zhu}\ \emph {et~al.}(1995{\natexlab{a}})\citenamefont
  {Zhu}, \citenamefont {Kleiman}, \citenamefont {Li}, \citenamefont {Lu},
  \citenamefont {Trentelman},\ and\ \citenamefont {Gordon}}]{Zhu1995}%
  \BibitemOpen
  \bibfield  {author} {\bibinfo {author} {\bibfnamefont {L.}~\bibnamefont
  {Zhu}}, \bibinfo {author} {\bibfnamefont {V.}~\bibnamefont {Kleiman}},
  \bibinfo {author} {\bibfnamefont {X.}~\bibnamefont {Li}}, \bibinfo {author}
  {\bibfnamefont {S.~P.}\ \bibnamefont {Lu}}, \bibinfo {author} {\bibfnamefont
  {K.}~\bibnamefont {Trentelman}}, \ and\ \bibinfo {author} {\bibfnamefont
  {R.~J.}\ \bibnamefont {Gordon}},\ }\href@noop {} {\bibfield  {journal}
  {\bibinfo  {journal} {Science}\ }\textbf {\bibinfo {volume} {270}},\ \bibinfo
  {pages} {77} (\bibinfo {year} {1995}{\natexlab{a}})}\BibitemShut {NoStop}%
\bibitem [{\citenamefont {Lavigne}\ and\ \citenamefont
  {Brumer}(2020)}]{Lavigne:20}%
  \BibitemOpen
  \bibfield  {author} {\bibinfo {author} {\bibfnamefont {C.}~\bibnamefont
  {Lavigne}}\ and\ \bibinfo {author} {\bibfnamefont {P.}~\bibnamefont
  {Brumer}},\ }\href {\doibase 10.1063/5.0012642} {\bibfield  {journal}
  {\bibinfo  {journal} {J. Chem. Phys.}\ }\textbf {\bibinfo {volume} {153}},\
  \bibinfo {pages} {034303} (\bibinfo {year} {2020})}\BibitemShut {NoStop}%
\bibitem [{\citenamefont {Kleiman}\ \emph {et~al.}(1995)\citenamefont
  {Kleiman}, \citenamefont {Zhu}, \citenamefont {Li},\ and\ \citenamefont
  {Gordon}}]{Kleiman1995a}%
  \BibitemOpen
  \bibfield  {author} {\bibinfo {author} {\bibfnamefont {V.~D.}\ \bibnamefont
  {Kleiman}}, \bibinfo {author} {\bibfnamefont {L.}~\bibnamefont {Zhu}},
  \bibinfo {author} {\bibfnamefont {X.}~\bibnamefont {Li}}, \ and\ \bibinfo
  {author} {\bibfnamefont {R.~J.}\ \bibnamefont {Gordon}},\ }\href@noop {}
  {\bibfield  {journal} {\bibinfo  {journal} {J. Chem. Phys.}\ }\textbf
  {\bibinfo {volume} {102}},\ \bibinfo {pages} {5863} (\bibinfo {year}
  {1995})}\BibitemShut {NoStop}%
\bibitem [{\citenamefont {Grinev}\ \emph {et~al.}(2015)\citenamefont {Grinev},
  \citenamefont {Shapiro},\ and\ \citenamefont {Brumer}}]{Grinev2015}%
  \BibitemOpen
  \bibfield  {author} {\bibinfo {author} {\bibfnamefont {T.}~\bibnamefont
  {Grinev}}, \bibinfo {author} {\bibfnamefont {M.}~\bibnamefont {Shapiro}}, \
  and\ \bibinfo {author} {\bibfnamefont {P.}~\bibnamefont {Brumer}},\ }\href
  {\doibase 10.1088/0953-4075/48/17/174004} {\bibfield  {journal} {\bibinfo
  {journal} {J. Phys. B}\ }\textbf {\bibinfo {volume} {48}},\ \bibinfo {pages}
  {174004} (\bibinfo {year} {2015})}\BibitemShut {NoStop}%
\bibitem [{\citenamefont {Brumer}\ and\ \citenamefont
  {Shapiro}(1986{\natexlab{a}})}]{Brumer1986a}%
  \BibitemOpen
  \bibfield  {author} {\bibinfo {author} {\bibfnamefont {P.}~\bibnamefont
  {Brumer}}\ and\ \bibinfo {author} {\bibfnamefont {M.}~\bibnamefont
  {Shapiro}},\ }\href@noop {} {\bibfield  {journal} {\bibinfo  {journal} {Chem.
  Phys. Lett.}\ }\textbf {\bibinfo {volume} {126}},\ \bibinfo {pages} {541}
  (\bibinfo {year} {1986}{\natexlab{a}})}\BibitemShut {NoStop}%
\bibitem [{\citenamefont {Brumer}\ and\ \citenamefont
  {Shapiro}(1986{\natexlab{b}})}]{Brumer1986b}%
  \BibitemOpen
  \bibfield  {author} {\bibinfo {author} {\bibfnamefont {P.}~\bibnamefont
  {Brumer}}\ and\ \bibinfo {author} {\bibfnamefont {M.}~\bibnamefont
  {Shapiro}},\ }\href@noop {} {\bibfield  {journal} {\bibinfo  {journal}
  {Faraday Discuss. Chem. Soc.}\ }\textbf {\bibinfo {volume} {82}},\ \bibinfo
  {pages} {177} (\bibinfo {year} {1986}{\natexlab{b}})}\BibitemShut {NoStop}%
\bibitem [{\citenamefont {Tannor}\ and\ \citenamefont
  {Rice}(1985)}]{Tannor1985}%
  \BibitemOpen
  \bibfield  {author} {\bibinfo {author} {\bibfnamefont {D.~J.}\ \bibnamefont
  {Tannor}}\ and\ \bibinfo {author} {\bibfnamefont {S.~A.}\ \bibnamefont
  {Rice}},\ }\href@noop {} {\bibfield  {journal} {\bibinfo  {journal} {J. Chem.
  Phys.}\ }\textbf {\bibinfo {volume} {83}},\ \bibinfo {pages} {5013} (\bibinfo
  {year} {1985})}\BibitemShut {NoStop}%
\bibitem [{\citenamefont {Sheehy}\ \emph
  {et~al.}(1995{\natexlab{b}})\citenamefont {Sheehy}, \citenamefont {Walker},\
  and\ \citenamefont {DiMauro}}]{Sheehy:95}%
  \BibitemOpen
  \bibfield  {author} {\bibinfo {author} {\bibfnamefont {B.}~\bibnamefont
  {Sheehy}}, \bibinfo {author} {\bibfnamefont {B.}~\bibnamefont {Walker}}, \
  and\ \bibinfo {author} {\bibfnamefont {L.~F.}\ \bibnamefont {DiMauro}},\
  }\href {\doibase 10.1103/PhysRevLett.74.4799} {\bibfield  {journal} {\bibinfo
   {journal} {Phys. Rev. Lett.}\ }\textbf {\bibinfo {volume} {74}},\ \bibinfo
  {pages} {4799} (\bibinfo {year} {1995}{\natexlab{b}})}\BibitemShut {NoStop}%
\bibitem [{\citenamefont {Zhu}\ \emph {et~al.}(1995{\natexlab{b}})\citenamefont
  {Zhu}, \citenamefont {Kleiman}, \citenamefont {Li}, \citenamefont {Lu},
  \citenamefont {Trentelman},\ and\ \citenamefont {Gordon}}]{Zhu:95}%
  \BibitemOpen
  \bibfield  {author} {\bibinfo {author} {\bibfnamefont {L.}~\bibnamefont
  {Zhu}}, \bibinfo {author} {\bibfnamefont {V.}~\bibnamefont {Kleiman}},
  \bibinfo {author} {\bibfnamefont {X.}~\bibnamefont {Li}}, \bibinfo {author}
  {\bibfnamefont {S.~P.}\ \bibnamefont {Lu}}, \bibinfo {author} {\bibfnamefont
  {K.}~\bibnamefont {Trentelman}}, \ and\ \bibinfo {author} {\bibfnamefont
  {R.~J.}\ \bibnamefont {Gordon}},\ }\href {\doibase
  10.1126/science.270.5233.77} {\bibfield  {journal} {\bibinfo  {journal}
  {Science}\ }\textbf {\bibinfo {volume} {270}},\ \bibinfo {pages} {77}
  (\bibinfo {year} {1995}{\natexlab{b}})}\BibitemShut {NoStop}%
\bibitem [{\citenamefont {Shnitman}\ \emph {et~al.}(1996)\citenamefont
  {Shnitman}, \citenamefont {Sofer}, \citenamefont {Golub}, \citenamefont
  {Yogev}, \citenamefont {Shapiro}, \citenamefont {Chen},\ and\ \citenamefont
  {Brumer}}]{Shnitman:96}%
  \BibitemOpen
  \bibfield  {author} {\bibinfo {author} {\bibfnamefont {A.}~\bibnamefont
  {Shnitman}}, \bibinfo {author} {\bibfnamefont {I.}~\bibnamefont {Sofer}},
  \bibinfo {author} {\bibfnamefont {I.}~\bibnamefont {Golub}}, \bibinfo
  {author} {\bibfnamefont {A.}~\bibnamefont {Yogev}}, \bibinfo {author}
  {\bibfnamefont {M.}~\bibnamefont {Shapiro}}, \bibinfo {author} {\bibfnamefont
  {Z.}~\bibnamefont {Chen}}, \ and\ \bibinfo {author} {\bibfnamefont
  {P.}~\bibnamefont {Brumer}},\ }\href {\doibase 10.1103/PhysRevLett.76.2886}
  {\bibfield  {journal} {\bibinfo  {journal} {Phys. Rev. Lett.}\ }\textbf
  {\bibinfo {volume} {76}},\ \bibinfo {pages} {2886} (\bibinfo {year}
  {1996})}\BibitemShut {NoStop}%
\bibitem [{\citenamefont {Shapiro}\ and\ \citenamefont
  {Brumer}(1996)}]{Shapiro1996}%
  \BibitemOpen
  \bibfield  {author} {\bibinfo {author} {\bibfnamefont {M.}~\bibnamefont
  {Shapiro}}\ and\ \bibinfo {author} {\bibfnamefont {P.}~\bibnamefont
  {Brumer}},\ }\href@noop {} {\bibfield  {journal} {\bibinfo  {journal} {Phys.
  Rev. Lett.}\ }\textbf {\bibinfo {volume} {77}},\ \bibinfo {pages} {2574}
  (\bibinfo {year} {1996})}\BibitemShut {NoStop}%
\bibitem [{\citenamefont {Omiste}\ \emph {et~al.}(2018)\citenamefont {Omiste},
  \citenamefont {Flo\ss},\ and\ \citenamefont {Brumer}}]{Omiste2018}%
  \BibitemOpen
  \bibfield  {author} {\bibinfo {author} {\bibfnamefont {J.~J.}\ \bibnamefont
  {Omiste}}, \bibinfo {author} {\bibfnamefont {J.}~\bibnamefont {Flo\ss}}, \
  and\ \bibinfo {author} {\bibfnamefont {P.}~\bibnamefont {Brumer}},\
  }\href@noop {} {\bibfield  {journal} {\bibinfo  {journal} {Phys. Rev. Lett.}\
  }\textbf {\bibinfo {volume} {121}},\ \bibinfo {pages} {163405} (\bibinfo
  {year} {2018})}\BibitemShut {NoStop}%
\bibitem [{\citenamefont {Krems}(2008)}]{Krems:08}%
  \BibitemOpen
  \bibfield  {author} {\bibinfo {author} {\bibfnamefont {R.~V.}\ \bibnamefont
  {Krems}},\ }\href@noop {} {\bibfield  {journal} {\bibinfo  {journal} {Phys.
  Chem. Chem. Phys.}\ }\textbf {\bibinfo {volume} {10}},\ \bibinfo {pages}
  {4079} (\bibinfo {year} {2008})}\BibitemShut {NoStop}%
\bibitem [{\citenamefont {Bohn}\ \emph {et~al.}(2017)\citenamefont {Bohn},
  \citenamefont {Rey},\ and\ \citenamefont {Ye}}]{Bohn:17}%
  \BibitemOpen
  \bibfield  {author} {\bibinfo {author} {\bibfnamefont {J.~L.}\ \bibnamefont
  {Bohn}}, \bibinfo {author} {\bibfnamefont {A.~M.}\ \bibnamefont {Rey}}, \
  and\ \bibinfo {author} {\bibfnamefont {J.}~\bibnamefont {Ye}},\ }\href@noop
  {} {\bibfield  {journal} {\bibinfo  {journal} {Science}\ }\textbf {\bibinfo
  {volume} {357}},\ \bibinfo {pages} {1002} (\bibinfo {year}
  {2017})}\BibitemShut {NoStop}%
\bibitem [{\citenamefont {Henson}\ \emph {et~al.}(2012)\citenamefont {Henson},
  \citenamefont {Gersten}, \citenamefont {Shagam}, \citenamefont {Narevicius},\
  and\ \citenamefont {Narevicius}}]{Henson:12}%
  \BibitemOpen
  \bibfield  {author} {\bibinfo {author} {\bibfnamefont {A.~B.}\ \bibnamefont
  {Henson}}, \bibinfo {author} {\bibfnamefont {S.}~\bibnamefont {Gersten}},
  \bibinfo {author} {\bibfnamefont {Y.}~\bibnamefont {Shagam}}, \bibinfo
  {author} {\bibfnamefont {J.}~\bibnamefont {Narevicius}}, \ and\ \bibinfo
  {author} {\bibfnamefont {E.}~\bibnamefont {Narevicius}},\ }\href {\doibase
  10.1126/science.1229141} {\bibfield  {journal} {\bibinfo  {journal}
  {Science}\ }\textbf {\bibinfo {volume} {338}},\ \bibinfo {pages} {234}
  (\bibinfo {year} {2012})}\BibitemShut {NoStop}%
\bibitem [{\citenamefont {Klein}\ \emph {et~al.}(2016)\citenamefont {Klein},
  \citenamefont {Shagam}, \citenamefont {Skomorowski}, \citenamefont
  {{\.Z}uchowski}, \citenamefont {Pawlak}, \citenamefont {Janssen},
  \citenamefont {Moiseyev}, \citenamefont {van~de Meerakker}, \citenamefont
  {van~der Avoird}, \citenamefont {Koch},\ and\ \citenamefont
  {Narevicius}}]{Klein:16}%
  \BibitemOpen
  \bibfield  {author} {\bibinfo {author} {\bibfnamefont {A.}~\bibnamefont
  {Klein}}, \bibinfo {author} {\bibfnamefont {Y.}~\bibnamefont {Shagam}},
  \bibinfo {author} {\bibfnamefont {W.}~\bibnamefont {Skomorowski}}, \bibinfo
  {author} {\bibfnamefont {P.~S.}\ \bibnamefont {{\.Z}uchowski}}, \bibinfo
  {author} {\bibfnamefont {M.}~\bibnamefont {Pawlak}}, \bibinfo {author}
  {\bibfnamefont {L.~M.~C.}\ \bibnamefont {Janssen}}, \bibinfo {author}
  {\bibfnamefont {N.}~\bibnamefont {Moiseyev}}, \bibinfo {author}
  {\bibfnamefont {S.~Y.~T.}\ \bibnamefont {van~de Meerakker}}, \bibinfo
  {author} {\bibfnamefont {A.}~\bibnamefont {van~der Avoird}}, \bibinfo
  {author} {\bibfnamefont {C.~P.}\ \bibnamefont {Koch}}, \ and\ \bibinfo
  {author} {\bibfnamefont {E.}~\bibnamefont {Narevicius}},\ }\href@noop {}
  {\bibfield  {journal} {\bibinfo  {journal} {Nat. Phys.}\ }\textbf {\bibinfo
  {volume} {13}},\ \bibinfo {pages} {35} (\bibinfo {year} {2016})}\BibitemShut
  {NoStop}%
\bibitem [{\citenamefont {Vogels}\ \emph {et~al.}(2015)\citenamefont {Vogels},
  \citenamefont {Onvlee}, \citenamefont {Chefdeville}, \citenamefont {van~der
  Avoird}, \citenamefont {Groenenboom},\ and\ \citenamefont {van~de
  Meerakker}}]{Vogels:15}%
  \BibitemOpen
  \bibfield  {author} {\bibinfo {author} {\bibfnamefont {S.~N.}\ \bibnamefont
  {Vogels}}, \bibinfo {author} {\bibfnamefont {J.}~\bibnamefont {Onvlee}},
  \bibinfo {author} {\bibfnamefont {S.}~\bibnamefont {Chefdeville}}, \bibinfo
  {author} {\bibfnamefont {A.}~\bibnamefont {van~der Avoird}}, \bibinfo
  {author} {\bibfnamefont {G.~C.}\ \bibnamefont {Groenenboom}}, \ and\ \bibinfo
  {author} {\bibfnamefont {S.~Y.~T.}\ \bibnamefont {van~de Meerakker}},\
  }\href@noop {} {\bibfield  {journal} {\bibinfo  {journal} {Science}\ }\textbf
  {\bibinfo {volume} {350}},\ \bibinfo {pages} {787} (\bibinfo {year}
  {2015})}\BibitemShut {NoStop}%
\bibitem [{\citenamefont {Perreault}\ \emph {et~al.}(2017)\citenamefont
  {Perreault}, \citenamefont {Mukherjee},\ and\ \citenamefont
  {Zare}}]{Perreault2017}%
  \BibitemOpen
  \bibfield  {author} {\bibinfo {author} {\bibfnamefont {W.~E.}\ \bibnamefont
  {Perreault}}, \bibinfo {author} {\bibfnamefont {N.}~\bibnamefont
  {Mukherjee}}, \ and\ \bibinfo {author} {\bibfnamefont {R.~N.}\ \bibnamefont
  {Zare}},\ }\href@noop {} {\bibfield  {journal} {\bibinfo  {journal}
  {Science}\ }\textbf {\bibinfo {volume} {358}},\ \bibinfo {pages} {356}
  (\bibinfo {year} {2017})}\BibitemShut {NoStop}%
\bibitem [{\citenamefont {Perreault}\ \emph {et~al.}(2018)\citenamefont
  {Perreault}, \citenamefont {Mukherjee},\ and\ \citenamefont
  {Zare}}]{Perreault:18}%
  \BibitemOpen
  \bibfield  {author} {\bibinfo {author} {\bibfnamefont {W.~E.}\ \bibnamefont
  {Perreault}}, \bibinfo {author} {\bibfnamefont {N.}~\bibnamefont
  {Mukherjee}}, \ and\ \bibinfo {author} {\bibfnamefont {R.~N.}\ \bibnamefont
  {Zare}},\ }\href {\doibase 10.1038/s41557-018-0028-5} {\bibfield  {journal}
  {\bibinfo  {journal} {Nat. Chem.}\ }\textbf {\bibinfo {volume} {10}},\
  \bibinfo {pages} {561} (\bibinfo {year} {2018})}\BibitemShut {NoStop}%
\bibitem [{\citenamefont {Tizniti}\ \emph {et~al.}(2014)\citenamefont
  {Tizniti}, \citenamefont {Picard}, \citenamefont {Lique}, \citenamefont
  {Berteloite}, \citenamefont {Canosa}, \citenamefont {Alexander},\ and\
  \citenamefont {Sims}}]{Tizniti2014}%
  \BibitemOpen
  \bibfield  {author} {\bibinfo {author} {\bibfnamefont {M.}~\bibnamefont
  {Tizniti}}, \bibinfo {author} {\bibfnamefont {S.~D.~L.}\ \bibnamefont
  {Picard}}, \bibinfo {author} {\bibfnamefont {F.}~\bibnamefont {Lique}},
  \bibinfo {author} {\bibfnamefont {C.}~\bibnamefont {Berteloite}}, \bibinfo
  {author} {\bibfnamefont {A.}~\bibnamefont {Canosa}}, \bibinfo {author}
  {\bibfnamefont {M.~H.}\ \bibnamefont {Alexander}}, \ and\ \bibinfo {author}
  {\bibfnamefont {I.~R.}\ \bibnamefont {Sims}},\ }\href@noop {} {\bibfield
  {journal} {\bibinfo  {journal} {Nat. Chem.}\ }\textbf {\bibinfo {volume}
  {6}},\ \bibinfo {pages} {141} (\bibinfo {year} {2014})}\BibitemShut {NoStop}%
\bibitem [{\citenamefont {Ni}\ \emph {et~al.}(2010)\citenamefont {Ni},
  \citenamefont {Ospelkaus}, \citenamefont {Wang}, \citenamefont
  {Qu{\'e}m{\'e}ner}, \citenamefont {Neyenhuis}, \citenamefont {de~Miranda},
  \citenamefont {Bohn}, \citenamefont {Ye},\ and\ \citenamefont {Jin}}]{Ni:10}%
  \BibitemOpen
  \bibfield  {author} {\bibinfo {author} {\bibfnamefont {K.-K.}\ \bibnamefont
  {Ni}}, \bibinfo {author} {\bibfnamefont {S.}~\bibnamefont {Ospelkaus}},
  \bibinfo {author} {\bibfnamefont {D.}~\bibnamefont {Wang}}, \bibinfo {author}
  {\bibfnamefont {G.}~\bibnamefont {Qu{\'e}m{\'e}ner}}, \bibinfo {author}
  {\bibfnamefont {B.}~\bibnamefont {Neyenhuis}}, \bibinfo {author}
  {\bibfnamefont {M.~H.~G.}\ \bibnamefont {de~Miranda}}, \bibinfo {author}
  {\bibfnamefont {J.~L.}\ \bibnamefont {Bohn}}, \bibinfo {author}
  {\bibfnamefont {J.}~\bibnamefont {Ye}}, \ and\ \bibinfo {author}
  {\bibfnamefont {D.~S.}\ \bibnamefont {Jin}},\ }\href@noop {} {\bibfield
  {journal} {\bibinfo  {journal} {Nature (London)}\ }\textbf {\bibinfo {volume}
  {464}},\ \bibinfo {pages} {1324} (\bibinfo {year} {2010})}\BibitemShut
  {NoStop}%
\bibitem [{\citenamefont {de~Miranda}\ \emph {et~al.}(2011)\citenamefont
  {de~Miranda}, \citenamefont {Chotia}, \citenamefont {Neyenhuis},
  \citenamefont {Wang}, \citenamefont {Qu{\'e}m{\'e}ner}, \citenamefont
  {Ospelkaus}, \citenamefont {Bohn}, \citenamefont {Ye},\ and\ \citenamefont
  {Jin}}]{Miranda:11}%
  \BibitemOpen
  \bibfield  {author} {\bibinfo {author} {\bibfnamefont {M.~H.~G.}\
  \bibnamefont {de~Miranda}}, \bibinfo {author} {\bibfnamefont
  {A.}~\bibnamefont {Chotia}}, \bibinfo {author} {\bibfnamefont
  {B.}~\bibnamefont {Neyenhuis}}, \bibinfo {author} {\bibfnamefont
  {D.}~\bibnamefont {Wang}}, \bibinfo {author} {\bibfnamefont {G.}~\bibnamefont
  {Qu{\'e}m{\'e}ner}}, \bibinfo {author} {\bibfnamefont {S.}~\bibnamefont
  {Ospelkaus}}, \bibinfo {author} {\bibfnamefont {J.~L.}\ \bibnamefont {Bohn}},
  \bibinfo {author} {\bibfnamefont {J.}~\bibnamefont {Ye}}, \ and\ \bibinfo
  {author} {\bibfnamefont {D.~S.}\ \bibnamefont {Jin}},\ }\href@noop {}
  {\bibfield  {journal} {\bibinfo  {journal} {Nat. Phys.}\ }\textbf {\bibinfo
  {volume} {7}},\ \bibinfo {pages} {502} (\bibinfo {year} {2011})}\BibitemShut
  {NoStop}%
\bibitem [{\citenamefont {Aldegunde}\ \emph {et~al.}(2006)\citenamefont
  {Aldegunde}, \citenamefont {Alvari{\~n}o}, \citenamefont {de~Miranda},
  \citenamefont {S{\'a}ez~R{\'a}banos},\ and\ \citenamefont
  {Aoiz}}]{Aldegunde:06}%
  \BibitemOpen
  \bibfield  {author} {\bibinfo {author} {\bibfnamefont {J.}~\bibnamefont
  {Aldegunde}}, \bibinfo {author} {\bibfnamefont {J.~M.}\ \bibnamefont
  {Alvari{\~n}o}}, \bibinfo {author} {\bibfnamefont {M.~P.}\ \bibnamefont
  {de~Miranda}}, \bibinfo {author} {\bibfnamefont {V.}~\bibnamefont
  {S{\'a}ez~R{\'a}banos}}, \ and\ \bibinfo {author} {\bibfnamefont {F.~J.}\
  \bibnamefont {Aoiz}},\ }\href {\doibase 10.1063/1.2212418} {\bibfield
  {journal} {\bibinfo  {journal} {J. Chem. Phys.}\ }\textbf {\bibinfo {volume}
  {125}},\ \bibinfo {pages} {133104} (\bibinfo {year} {2006})}\BibitemShut
  {NoStop}%
\bibitem [{\citenamefont {Croft}\ \emph {et~al.}(2018)\citenamefont {Croft},
  \citenamefont {Balakrishnan}, \citenamefont {Huang},\ and\ \citenamefont
  {Guo}}]{Croft:18}%
  \BibitemOpen
  \bibfield  {author} {\bibinfo {author} {\bibfnamefont {J.~F.~E.}\
  \bibnamefont {Croft}}, \bibinfo {author} {\bibfnamefont {N.}~\bibnamefont
  {Balakrishnan}}, \bibinfo {author} {\bibfnamefont {M.}~\bibnamefont {Huang}},
  \ and\ \bibinfo {author} {\bibfnamefont {H.}~\bibnamefont {Guo}},\ }\href
  {\doibase 10.1103/PhysRevLett.121.113401} {\bibfield  {journal} {\bibinfo
  {journal} {Phys. Rev. Lett.}\ }\textbf {\bibinfo {volume} {121}},\ \bibinfo
  {pages} {113401} (\bibinfo {year} {2018})}\BibitemShut {NoStop}%
\bibitem [{\citenamefont {Jambrina}\ \emph {et~al.}(2019)\citenamefont
  {Jambrina}, \citenamefont {Croft}, \citenamefont {Guo}, \citenamefont
  {Brouard}, \citenamefont {Balakrishnan},\ and\ \citenamefont
  {Aoiz}}]{Jambrina:19}%
  \BibitemOpen
  \bibfield  {author} {\bibinfo {author} {\bibfnamefont {P.~G.}\ \bibnamefont
  {Jambrina}}, \bibinfo {author} {\bibfnamefont {J.~F.~E.}\ \bibnamefont
  {Croft}}, \bibinfo {author} {\bibfnamefont {H.}~\bibnamefont {Guo}}, \bibinfo
  {author} {\bibfnamefont {M.}~\bibnamefont {Brouard}}, \bibinfo {author}
  {\bibfnamefont {N.}~\bibnamefont {Balakrishnan}}, \ and\ \bibinfo {author}
  {\bibfnamefont {F.~J.}\ \bibnamefont {Aoiz}},\ }\href {\doibase
  10.1103/PhysRevLett.123.043401} {\bibfield  {journal} {\bibinfo  {journal}
  {Phys. Rev. Lett.}\ }\textbf {\bibinfo {volume} {123}},\ \bibinfo {pages}
  {043401} (\bibinfo {year} {2019})}\BibitemShut {NoStop}%
\bibitem [{\citenamefont {Chin}\ \emph {et~al.}(2010)\citenamefont {Chin},
  \citenamefont {Grimm}, \citenamefont {Julienne},\ and\ \citenamefont
  {Tiesinga}}]{Chin:10}%
  \BibitemOpen
  \bibfield  {author} {\bibinfo {author} {\bibfnamefont {C.}~\bibnamefont
  {Chin}}, \bibinfo {author} {\bibfnamefont {R.}~\bibnamefont {Grimm}},
  \bibinfo {author} {\bibfnamefont {P.}~\bibnamefont {Julienne}}, \ and\
  \bibinfo {author} {\bibfnamefont {E.}~\bibnamefont {Tiesinga}},\ }\href
  {\doibase 10.1103/RevModPhys.82.1225} {\bibfield  {journal} {\bibinfo
  {journal} {Rev. Mod. Phys.}\ }\textbf {\bibinfo {volume} {82}},\ \bibinfo
  {pages} {1225} (\bibinfo {year} {2010})}\BibitemShut {NoStop}%
\bibitem [{\citenamefont {Derevianko}\ and\ \citenamefont
  {Katori}(2011)}]{Derevianko:11}%
  \BibitemOpen
  \bibfield  {author} {\bibinfo {author} {\bibfnamefont {A.}~\bibnamefont
  {Derevianko}}\ and\ \bibinfo {author} {\bibfnamefont {H.}~\bibnamefont
  {Katori}},\ }\href {\doibase 10.1103/RevModPhys.83.331} {\bibfield  {journal}
  {\bibinfo  {journal} {Rev. Mod. Phys.}\ }\textbf {\bibinfo {volume} {83}},\
  \bibinfo {pages} {331} (\bibinfo {year} {2011})}\BibitemShut {NoStop}%
\bibitem [{\citenamefont {Ludlow}\ \emph {et~al.}(2015)\citenamefont {Ludlow},
  \citenamefont {Boyd}, \citenamefont {Ye}, \citenamefont {Peik},\ and\
  \citenamefont {Schmidt}}]{Ludlow:15}%
  \BibitemOpen
  \bibfield  {author} {\bibinfo {author} {\bibfnamefont {A.~D.}\ \bibnamefont
  {Ludlow}}, \bibinfo {author} {\bibfnamefont {M.~M.}\ \bibnamefont {Boyd}},
  \bibinfo {author} {\bibfnamefont {J.}~\bibnamefont {Ye}}, \bibinfo {author}
  {\bibfnamefont {E.}~\bibnamefont {Peik}}, \ and\ \bibinfo {author}
  {\bibfnamefont {P.~O.}\ \bibnamefont {Schmidt}},\ }\href {\doibase
  10.1103/RevModPhys.87.637} {\bibfield  {journal} {\bibinfo  {journal} {Rev.
  Mod. Phys.}\ }\textbf {\bibinfo {volume} {87}},\ \bibinfo {pages} {637}
  (\bibinfo {year} {2015})}\BibitemShut {NoStop}%
\bibitem [{\citenamefont {Ospelkaus}\ \emph {et~al.}(2010)\citenamefont
  {Ospelkaus}, \citenamefont {Ni}, \citenamefont {Wang}, \citenamefont
  {de~Miranda}, \citenamefont {Neyenhuis}, \citenamefont {Qu{\'e}m{\'e}ner},
  \citenamefont {Julienne}, \citenamefont {Bohn}, \citenamefont {Jin},\ and\
  \citenamefont {Ye}}]{Ospelkaus:10}%
  \BibitemOpen
  \bibfield  {author} {\bibinfo {author} {\bibfnamefont {S.}~\bibnamefont
  {Ospelkaus}}, \bibinfo {author} {\bibfnamefont {K.-K.}\ \bibnamefont {Ni}},
  \bibinfo {author} {\bibfnamefont {D.}~\bibnamefont {Wang}}, \bibinfo {author}
  {\bibfnamefont {M.~H.~G.}\ \bibnamefont {de~Miranda}}, \bibinfo {author}
  {\bibfnamefont {B.}~\bibnamefont {Neyenhuis}}, \bibinfo {author}
  {\bibfnamefont {G.}~\bibnamefont {Qu{\'e}m{\'e}ner}}, \bibinfo {author}
  {\bibfnamefont {P.~S.}\ \bibnamefont {Julienne}}, \bibinfo {author}
  {\bibfnamefont {J.~L.}\ \bibnamefont {Bohn}}, \bibinfo {author}
  {\bibfnamefont {D.~S.}\ \bibnamefont {Jin}}, \ and\ \bibinfo {author}
  {\bibfnamefont {J.}~\bibnamefont {Ye}},\ }\href@noop {} {\bibfield  {journal}
  {\bibinfo  {journal} {Science}\ }\textbf {\bibinfo {volume} {327}},\ \bibinfo
  {pages} {853} (\bibinfo {year} {2010})}\BibitemShut {NoStop}%
\bibitem [{\citenamefont {Park}\ \emph {et~al.}(2015)\citenamefont {Park},
  \citenamefont {Will},\ and\ \citenamefont {Zwierlein}}]{Park:15}%
  \BibitemOpen
  \bibfield  {author} {\bibinfo {author} {\bibfnamefont {J.~W.}\ \bibnamefont
  {Park}}, \bibinfo {author} {\bibfnamefont {S.~A.}\ \bibnamefont {Will}}, \
  and\ \bibinfo {author} {\bibfnamefont {M.~W.}\ \bibnamefont {Zwierlein}},\
  }\href {\doibase 10.1103/PhysRevLett.114.205302} {\bibfield  {journal}
  {\bibinfo  {journal} {Phys. Rev. Lett.}\ }\textbf {\bibinfo {volume} {114}},\
  \bibinfo {pages} {205302} (\bibinfo {year} {2015})}\BibitemShut {NoStop}%
\bibitem [{\citenamefont {Rvachov}\ \emph {et~al.}(2017)\citenamefont
  {Rvachov}, \citenamefont {Son}, \citenamefont {Sommer}, \citenamefont
  {Ebadi}, \citenamefont {Park}, \citenamefont {Zwierlein}, \citenamefont
  {Ketterle},\ and\ \citenamefont {Jamison}}]{Rvachov:17}%
  \BibitemOpen
  \bibfield  {author} {\bibinfo {author} {\bibfnamefont {T.~M.}\ \bibnamefont
  {Rvachov}}, \bibinfo {author} {\bibfnamefont {H.}~\bibnamefont {Son}},
  \bibinfo {author} {\bibfnamefont {A.~T.}\ \bibnamefont {Sommer}}, \bibinfo
  {author} {\bibfnamefont {S.}~\bibnamefont {Ebadi}}, \bibinfo {author}
  {\bibfnamefont {J.~J.}\ \bibnamefont {Park}}, \bibinfo {author}
  {\bibfnamefont {M.~W.}\ \bibnamefont {Zwierlein}}, \bibinfo {author}
  {\bibfnamefont {W.}~\bibnamefont {Ketterle}}, \ and\ \bibinfo {author}
  {\bibfnamefont {A.~O.}\ \bibnamefont {Jamison}},\ }\href {\doibase
  10.1103/PhysRevLett.119.143001} {\bibfield  {journal} {\bibinfo  {journal}
  {Phys. Rev. Lett.}\ }\textbf {\bibinfo {volume} {119}},\ \bibinfo {pages}
  {143001} (\bibinfo {year} {2017})}\BibitemShut {NoStop}%
\bibitem [{\citenamefont {Barry}\ \emph {et~al.}(2014)\citenamefont {Barry},
  \citenamefont {McCarron}, \citenamefont {Norrgard}, \citenamefont
  {Steinecker},\ and\ \citenamefont {DeMille}}]{Barry:14}%
  \BibitemOpen
  \bibfield  {author} {\bibinfo {author} {\bibfnamefont {J.~F.}\ \bibnamefont
  {Barry}}, \bibinfo {author} {\bibfnamefont {D.~J.}\ \bibnamefont {McCarron}},
  \bibinfo {author} {\bibfnamefont {E.~B.}\ \bibnamefont {Norrgard}}, \bibinfo
  {author} {\bibfnamefont {M.~H.}\ \bibnamefont {Steinecker}}, \ and\ \bibinfo
  {author} {\bibfnamefont {D.}~\bibnamefont {DeMille}},\ }\href@noop {}
  {\bibfield  {journal} {\bibinfo  {journal} {Nature (London)}\ }\textbf
  {\bibinfo {volume} {512}},\ \bibinfo {pages} {286} (\bibinfo {year}
  {2014})}\BibitemShut {NoStop}%
\bibitem [{\citenamefont {McCarron}\ \emph {et~al.}(2018)\citenamefont
  {McCarron}, \citenamefont {Steinecker}, \citenamefont {Zhu},\ and\
  \citenamefont {DeMille}}]{McCarron:18}%
  \BibitemOpen
  \bibfield  {author} {\bibinfo {author} {\bibfnamefont {D.~J.}\ \bibnamefont
  {McCarron}}, \bibinfo {author} {\bibfnamefont {M.~H.}\ \bibnamefont
  {Steinecker}}, \bibinfo {author} {\bibfnamefont {Y.}~\bibnamefont {Zhu}}, \
  and\ \bibinfo {author} {\bibfnamefont {D.}~\bibnamefont {DeMille}},\ }\href
  {\doibase 10.1103/PhysRevLett.121.013202} {\bibfield  {journal} {\bibinfo
  {journal} {Phys. Rev. Lett.}\ }\textbf {\bibinfo {volume} {121}},\ \bibinfo
  {pages} {013202} (\bibinfo {year} {2018})}\BibitemShut {NoStop}%
\bibitem [{\citenamefont {Cheuk}\ \emph {et~al.}(2018)\citenamefont {Cheuk},
  \citenamefont {Anderegg}, \citenamefont {Augenbraun}, \citenamefont {Bao},
  \citenamefont {Burchesky}, \citenamefont {Ketterle},\ and\ \citenamefont
  {Doyle}}]{Cheuk:18}%
  \BibitemOpen
  \bibfield  {author} {\bibinfo {author} {\bibfnamefont {L.~W.}\ \bibnamefont
  {Cheuk}}, \bibinfo {author} {\bibfnamefont {L.}~\bibnamefont {Anderegg}},
  \bibinfo {author} {\bibfnamefont {B.~L.}\ \bibnamefont {Augenbraun}},
  \bibinfo {author} {\bibfnamefont {Y.}~\bibnamefont {Bao}}, \bibinfo {author}
  {\bibfnamefont {S.}~\bibnamefont {Burchesky}}, \bibinfo {author}
  {\bibfnamefont {W.}~\bibnamefont {Ketterle}}, \ and\ \bibinfo {author}
  {\bibfnamefont {J.~M.}\ \bibnamefont {Doyle}},\ }\href {\doibase
  10.1103/PhysRevLett.121.083201} {\bibfield  {journal} {\bibinfo  {journal}
  {Phys. Rev. Lett.}\ }\textbf {\bibinfo {volume} {121}},\ \bibinfo {pages}
  {083201} (\bibinfo {year} {2018})}\BibitemShut {NoStop}%
\bibitem [{\citenamefont {Anderegg}\ \emph {et~al.}(2018)\citenamefont
  {Anderegg}, \citenamefont {Augenbraun}, \citenamefont {Bao}, \citenamefont
  {Burchesky}, \citenamefont {Cheuk}, \citenamefont {Ketterle},\ and\
  \citenamefont {Doyle}}]{Anderegg:18}%
  \BibitemOpen
  \bibfield  {author} {\bibinfo {author} {\bibfnamefont {L.}~\bibnamefont
  {Anderegg}}, \bibinfo {author} {\bibfnamefont {B.~L.}\ \bibnamefont
  {Augenbraun}}, \bibinfo {author} {\bibfnamefont {Y.}~\bibnamefont {Bao}},
  \bibinfo {author} {\bibfnamefont {S.}~\bibnamefont {Burchesky}}, \bibinfo
  {author} {\bibfnamefont {L.~W.}\ \bibnamefont {Cheuk}}, \bibinfo {author}
  {\bibfnamefont {W.}~\bibnamefont {Ketterle}}, \ and\ \bibinfo {author}
  {\bibfnamefont {J.~M.}\ \bibnamefont {Doyle}},\ }\href {\doibase
  10.1038/s41567-018-0191-z} {\bibfield  {journal} {\bibinfo  {journal}
  {Nat.Phys.}\ }\textbf {\bibinfo {volume} {14}},\ \bibinfo {pages} {890}
  (\bibinfo {year} {2018})}\BibitemShut {NoStop}%
\bibitem [{\citenamefont {Kozyryev}\ \emph {et~al.}(2017)\citenamefont
  {Kozyryev}, \citenamefont {Baum}, \citenamefont {Matsuda}, \citenamefont
  {Augenbraun}, \citenamefont {Anderegg}, \citenamefont {Sedlack},\ and\
  \citenamefont {Doyle}}]{Kozyryev:17}%
  \BibitemOpen
  \bibfield  {author} {\bibinfo {author} {\bibfnamefont {I.}~\bibnamefont
  {Kozyryev}}, \bibinfo {author} {\bibfnamefont {L.}~\bibnamefont {Baum}},
  \bibinfo {author} {\bibfnamefont {K.}~\bibnamefont {Matsuda}}, \bibinfo
  {author} {\bibfnamefont {B.~L.}\ \bibnamefont {Augenbraun}}, \bibinfo
  {author} {\bibfnamefont {L.}~\bibnamefont {Anderegg}}, \bibinfo {author}
  {\bibfnamefont {A.~P.}\ \bibnamefont {Sedlack}}, \ and\ \bibinfo {author}
  {\bibfnamefont {J.~M.}\ \bibnamefont {Doyle}},\ }\href {\doibase
  10.1103/PhysRevLett.118.173201} {\bibfield  {journal} {\bibinfo  {journal}
  {Phys. Rev. Lett.}\ }\textbf {\bibinfo {volume} {118}},\ \bibinfo {pages}
  {173201} (\bibinfo {year} {2017})}\BibitemShut {NoStop}%
\bibitem [{\citenamefont {Mukherjee}\ and\ \citenamefont
  {Zare}(2011)}]{Mukherjee2011}%
  \BibitemOpen
  \bibfield  {author} {\bibinfo {author} {\bibfnamefont {N.}~\bibnamefont
  {Mukherjee}}\ and\ \bibinfo {author} {\bibfnamefont {R.}~\bibnamefont
  {Zare}},\ }\href@noop {} {\bibfield  {journal} {\bibinfo  {journal} {J. Chem.
  Phys.}\ }\textbf {\bibinfo {volume} {135}},\ \bibinfo {pages} {024201}
  (\bibinfo {year} {2011})}\BibitemShut {NoStop}%
\bibitem [{\citenamefont {Mukherjee}\ \emph {et~al.}(2014)\citenamefont
  {Mukherjee}, \citenamefont {Dong},\ and\ \citenamefont
  {Zare}}]{Mukherjee2014}%
  \BibitemOpen
  \bibfield  {author} {\bibinfo {author} {\bibfnamefont {N.}~\bibnamefont
  {Mukherjee}}, \bibinfo {author} {\bibfnamefont {W.}~\bibnamefont {Dong}}, \
  and\ \bibinfo {author} {\bibfnamefont {R.~N.}\ \bibnamefont {Zare}},\
  }\href@noop {} {\bibfield  {journal} {\bibinfo  {journal} {J. Chem. Phys.}\
  }\textbf {\bibinfo {volume} {140}},\ \bibinfo {pages} {074201} (\bibinfo
  {year} {2014})}\BibitemShut {NoStop}%
\bibitem [{\citenamefont {Scholak}\ and\ \citenamefont
  {Brumer}(2017)}]{Scholak:17}%
  \BibitemOpen
  \bibfield  {author} {\bibinfo {author} {\bibfnamefont {T.}~\bibnamefont
  {Scholak}}\ and\ \bibinfo {author} {\bibfnamefont {P.}~\bibnamefont
  {Brumer}},\ }\href@noop {} {\bibfield  {journal} {\bibinfo  {journal} {Adv.
  Chem. Phys.}\ }\textbf {\bibinfo {volume} {162}},\ \bibinfo {pages} {39}
  (\bibinfo {year} {2017})}\BibitemShut {NoStop}%
\bibitem [{\citenamefont {Qureshi}(2019)}]{Qureshi:19}%
  \BibitemOpen
  \bibfield  {author} {\bibinfo {author} {\bibfnamefont {T.}~\bibnamefont
  {Qureshi}},\ }\href@noop {} {\bibfield  {journal} {\bibinfo  {journal}
  {Quanta}\ }\textbf {\bibinfo {volume} {8}},\ \bibinfo {pages} {24} (\bibinfo
  {year} {2019})}\BibitemShut {NoStop}%
\bibitem [{\citenamefont {Skouteris}\ \emph {et~al.}(2000)\citenamefont
  {Skouteris}, \citenamefont {Castillo},\ and\ \citenamefont
  {Manolopoulos}}]{Skouteris2000}%
  \BibitemOpen
  \bibfield  {author} {\bibinfo {author} {\bibfnamefont {D.}~\bibnamefont
  {Skouteris}}, \bibinfo {author} {\bibfnamefont {J.~F.}\ \bibnamefont
  {Castillo}}, \ and\ \bibinfo {author} {\bibfnamefont {D.~E.}\ \bibnamefont
  {Manolopoulos}},\ }\href@noop {} {\bibfield  {journal} {\bibinfo  {journal}
  {Comp. Phys. Comm.}\ }\textbf {\bibinfo {volume} {133}},\ \bibinfo {pages}
  {128} (\bibinfo {year} {2000})}\BibitemShut {NoStop}%
\bibitem [{SM()}]{SM}%
  \BibitemOpen
  \href@noop {} {}\bibinfo {note} {See Supplemental Material at
  [http://link.aps.org/ supplemental/] for details of numerical calculations
  and convergence tests.}\BibitemShut {Stop}%
\bibitem [{\citenamefont {Balakrishnan}\ and\ \citenamefont
  {Dalgarno}(2001)}]{Balakrishnan2001}%
  \BibitemOpen
  \bibfield  {author} {\bibinfo {author} {\bibfnamefont {N.}~\bibnamefont
  {Balakrishnan}}\ and\ \bibinfo {author} {\bibfnamefont {A.}~\bibnamefont
  {Dalgarno}},\ }\href@noop {} {\bibfield  {journal} {\bibinfo  {journal}
  {Chem. Phys. Lett.}\ }\textbf {\bibinfo {volume} {341}},\ \bibinfo {pages}
  {652} (\bibinfo {year} {2001})}\BibitemShut {NoStop}%
\bibitem [{\citenamefont {Zhu}\ \emph {et~al.}(2002)\citenamefont {Zhu},
  \citenamefont {Krems}, \citenamefont {Dalgarno},\ and\ \citenamefont
  {Balakrishnan}}]{Zhu:02}%
  \BibitemOpen
  \bibfield  {author} {\bibinfo {author} {\bibfnamefont {C.}~\bibnamefont
  {Zhu}}, \bibinfo {author} {\bibfnamefont {R.}~\bibnamefont {Krems}}, \bibinfo
  {author} {\bibfnamefont {A.}~\bibnamefont {Dalgarno}}, \ and\ \bibinfo
  {author} {\bibfnamefont {N.}~\bibnamefont {Balakrishnan}},\ }\href {\doibase
  10.1086/342240} {\bibfield  {journal} {\bibinfo  {journal} {Astrophys. J.}\
  }\textbf {\bibinfo {volume} {577}},\ \bibinfo {pages} {795} (\bibinfo {year}
  {2002})}\BibitemShut {NoStop}%
\bibitem [{\citenamefont {Staanum}\ \emph {et~al.}(2006)\citenamefont
  {Staanum}, \citenamefont {Kraft}, \citenamefont {Lange}, \citenamefont
  {Wester},\ and\ \citenamefont {Weidem\"uller}}]{Staanum:06}%
  \BibitemOpen
  \bibfield  {author} {\bibinfo {author} {\bibfnamefont {P.}~\bibnamefont
  {Staanum}}, \bibinfo {author} {\bibfnamefont {S.~D.}\ \bibnamefont {Kraft}},
  \bibinfo {author} {\bibfnamefont {J.}~\bibnamefont {Lange}}, \bibinfo
  {author} {\bibfnamefont {R.}~\bibnamefont {Wester}}, \ and\ \bibinfo {author}
  {\bibfnamefont {M.}~\bibnamefont {Weidem\"uller}},\ }\href {\doibase
  10.1103/PhysRevLett.96.023201} {\bibfield  {journal} {\bibinfo  {journal}
  {Phys. Rev. Lett.}\ }\textbf {\bibinfo {volume} {96}},\ \bibinfo {pages}
  {023201} (\bibinfo {year} {2006})}\BibitemShut {NoStop}%
\bibitem [{\citenamefont {Zahzam}\ \emph {et~al.}(2006)\citenamefont {Zahzam},
  \citenamefont {Vogt}, \citenamefont {Mudrich}, \citenamefont {Comparat},\
  and\ \citenamefont {Pillet}}]{Zahzam:06}%
  \BibitemOpen
  \bibfield  {author} {\bibinfo {author} {\bibfnamefont {N.}~\bibnamefont
  {Zahzam}}, \bibinfo {author} {\bibfnamefont {T.}~\bibnamefont {Vogt}},
  \bibinfo {author} {\bibfnamefont {M.}~\bibnamefont {Mudrich}}, \bibinfo
  {author} {\bibfnamefont {D.}~\bibnamefont {Comparat}}, \ and\ \bibinfo
  {author} {\bibfnamefont {P.}~\bibnamefont {Pillet}},\ }\href {\doibase
  10.1103/PhysRevLett.96.023202} {\bibfield  {journal} {\bibinfo  {journal}
  {Phys. Rev. Lett.}\ }\textbf {\bibinfo {volume} {96}},\ \bibinfo {pages}
  {023202} (\bibinfo {year} {2006})}\BibitemShut {NoStop}%
\bibitem [{\citenamefont {Viteau}\ \emph {et~al.}(2008)\citenamefont {Viteau},
  \citenamefont {Chotia}, \citenamefont {Allegrini}, \citenamefont {Bouloufa},
  \citenamefont {Dulieu}, \citenamefont {Comparat},\ and\ \citenamefont
  {Pillet}}]{Viteau:08}%
  \BibitemOpen
  \bibfield  {author} {\bibinfo {author} {\bibfnamefont {M.}~\bibnamefont
  {Viteau}}, \bibinfo {author} {\bibfnamefont {A.}~\bibnamefont {Chotia}},
  \bibinfo {author} {\bibfnamefont {M.}~\bibnamefont {Allegrini}}, \bibinfo
  {author} {\bibfnamefont {N.}~\bibnamefont {Bouloufa}}, \bibinfo {author}
  {\bibfnamefont {O.}~\bibnamefont {Dulieu}}, \bibinfo {author} {\bibfnamefont
  {D.}~\bibnamefont {Comparat}}, \ and\ \bibinfo {author} {\bibfnamefont
  {P.}~\bibnamefont {Pillet}},\ }\href {\doibase 10.1126/science.1159496}
  {\bibfield  {journal} {\bibinfo  {journal} {Science}\ }\textbf {\bibinfo
  {volume} {321}},\ \bibinfo {pages} {232} (\bibinfo {year}
  {2008})}\BibitemShut {NoStop}%
\bibitem [{\citenamefont {Qu{\'e}m{\'e}ner}\ and\ \citenamefont
  {Julienne}(2012)}]{Quemener:12}%
  \BibitemOpen
  \bibfield  {author} {\bibinfo {author} {\bibfnamefont {G.}~\bibnamefont
  {Qu{\'e}m{\'e}ner}}\ and\ \bibinfo {author} {\bibfnamefont {P.~S.}\
  \bibnamefont {Julienne}},\ }\href {\doibase 10.1021/cr300092g} {\bibfield
  {journal} {\bibinfo  {journal} {Chemical Reviews}\ }\textbf {\bibinfo
  {volume} {112}},\ \bibinfo {pages} {4949} (\bibinfo {year}
  {2012})}\BibitemShut {NoStop}%
\bibitem [{\citenamefont {Althorpe}\ and\ \citenamefont
  {Clary}(2003)}]{Althorpe:03}%
  \BibitemOpen
  \bibfield  {author} {\bibinfo {author} {\bibfnamefont {S.~C.}\ \bibnamefont
  {Althorpe}}\ and\ \bibinfo {author} {\bibfnamefont {D.~C.}\ \bibnamefont
  {Clary}},\ }\href {\doibase 10.1146/annurev.physchem.54.011002.103750}
  {\bibfield  {journal} {\bibinfo  {journal} {Annu. Rev. Phys. Chem.}\ }\textbf
  {\bibinfo {volume} {54}},\ \bibinfo {pages} {493} (\bibinfo {year}
  {2003})}\BibitemShut {NoStop}%
\bibitem [{\citenamefont {Ren}\ \emph {et~al.}(2008)\citenamefont {Ren},
  \citenamefont {Che}, \citenamefont {Qiu}, \citenamefont {Wang}, \citenamefont
  {Dong}, \citenamefont {Dai}, \citenamefont {Wang}, \citenamefont {Yang},
  \citenamefont {Sun}, \citenamefont {Fu}, \citenamefont {Lee}, \citenamefont
  {Xu},\ and\ \citenamefont {Zhang}}]{Ren:08}%
  \BibitemOpen
  \bibfield  {author} {\bibinfo {author} {\bibfnamefont {Z.}~\bibnamefont
  {Ren}}, \bibinfo {author} {\bibfnamefont {L.}~\bibnamefont {Che}}, \bibinfo
  {author} {\bibfnamefont {M.}~\bibnamefont {Qiu}}, \bibinfo {author}
  {\bibfnamefont {X.}~\bibnamefont {Wang}}, \bibinfo {author} {\bibfnamefont
  {W.}~\bibnamefont {Dong}}, \bibinfo {author} {\bibfnamefont {D.}~\bibnamefont
  {Dai}}, \bibinfo {author} {\bibfnamefont {X.}~\bibnamefont {Wang}}, \bibinfo
  {author} {\bibfnamefont {X.}~\bibnamefont {Yang}}, \bibinfo {author}
  {\bibfnamefont {Z.}~\bibnamefont {Sun}}, \bibinfo {author} {\bibfnamefont
  {B.}~\bibnamefont {Fu}}, \bibinfo {author} {\bibfnamefont {S.-Y.}\
  \bibnamefont {Lee}}, \bibinfo {author} {\bibfnamefont {X.}~\bibnamefont
  {Xu}}, \ and\ \bibinfo {author} {\bibfnamefont {D.~H.}\ \bibnamefont
  {Zhang}},\ }\href {\doibase 10.1073/pnas.0709974105} {\bibfield  {journal}
  {\bibinfo  {journal} {Proc. Natl. Acad. Sci. USA}\ }\textbf {\bibinfo
  {volume} {105}},\ \bibinfo {pages} {12662} (\bibinfo {year}
  {2008})}\BibitemShut {NoStop}%
\bibitem [{\citenamefont {Zhang}\ and\ \citenamefont {Guo}(2016)}]{Zhang:16}%
  \BibitemOpen
  \bibfield  {author} {\bibinfo {author} {\bibfnamefont {D.~H.}\ \bibnamefont
  {Zhang}}\ and\ \bibinfo {author} {\bibfnamefont {H.}~\bibnamefont {Guo}},\
  }\href {\doibase 10.1146/annurev-physchem-040215-112016} {\bibfield
  {journal} {\bibinfo  {journal} {Annu. Rev. Phys. Chem.}\ }\textbf {\bibinfo
  {volume} {67}},\ \bibinfo {pages} {135} (\bibinfo {year} {2016})}\BibitemShut
  {NoStop}%
\end{thebibliography}
\end{document}